# The Affinity Effects of Parallelized Libraries in Concurrent Environments

FABIO LICHT, BRUNO SCHULZE, LUIS E. BONA, AND ANTONIO R. MURY[1]

Federal University of Parana (UFPR)

`licht@lncc.br`

## Abstract

The use of cloud computing grows as it appears to be an additional resource for High-Performance Parallel and Distributed Computing (HPDC), especially with respect to its use in support of scientific applications. Many studies have been devoted to determining the effect of the virtualization layer on the performance, but most of the studies conducted so far lack insight into the joint effects between application type, virtualization layer and parallelized libraries in applications.

This work introduces the concept of affinity with regard to the combined effects of the virtualization layer, class of application and parallelized libraries used in these applications. Affinity is here defined as the degree of influence that one application has on other applications when running concurrently in virtual environments hosted on the same real server.

The results presented here show how parallel libraries used in application implementation have a significant influence and how the combinations between these types of libraries and classes of applications could significantly influence the performance of the environment. In this context, the concept of affinity is then used to evaluate these impacts to contribute to better stability and performance in the computational environment.

## 1 Introduction

Growing concern about the quality of services provided by cloud computing has led researchers to seek mechanisms and methodologies to analyze and assist scheduling and the allocation of applications on computational resources. In this sense, to know how those applications interact when hosted in virtualized environments and the effects caused by concurrency on the real resources can contribute to minimizing the performance losses and instability among those environments. In a cloud computing environment, a single physical resource contains multiple virtualized environments where just one process, running in one of these virtual environments, could cause the degradation of the whole physical resource and consequently could impact all other existing environments on the same host.

In the specific case of cloud computing and its use as a support for high-performance parallel and distributed computing, especially from the perspective of scientific applications, studies are devoted to ascertaining the effect of the virtualization layer on the performance of those types of applications with different requirements compared to business support applications. The studies performed usually evaluate the loss of performance between clouds and do not address the effects of concurrency among different virtualized environments hosted on the same real cloud resource [1] [2] [4] [15].

The present work is part of an ongoing research program, and this paper will focus on evaluation of the effects of the concurrency caused by different classes of applications and the types of parallel libraries used to implement these applications. To date (to the best of our knowledge), there have been no studies that evaluate those combined effects. In this paper, we propose the concept of affinity, characterized by the degree of compatibility between classes of applications where concurrent execution in the same computing environment would result in a minimum loss for these applications and the environment itself. The closest concept to that proposed in this paper was presented by Ravi et al. [5], whose framework allows applications to run transparently inside virtual machines in one or more GPUs, seeking efficient virtualization of these GPUs and providing an indication of potential performance improvements in the consolidation process of the kernel.

To be able to validate the concept of affinity presented in this work, we used the application class approach. There are already studies that seek to categorize applications into classes, grouping these applications based on their characteristics in terms of use of computational resources. This type of categorization allows newly developed applications to be associated with one of these classes, bringing benefits by raising the possibility of being able to predict the behavior of these applications when run in a virtualized environment. In the tests, we used the Dwarf approach [6]. From the thirteen Dwarf classes proposed, we chose three classes of applications and four algorithms for this study. In one of the classes, we chose two algorithms to be used to verify different domains within the same class. The classes chosen represent the following types of applications: Dense Linear Algebra, Structured Grid, and Graph Transversal.

For the verification of the affinity level between classes, the tests analyzed the effects that the use of concurrently virtualized environments, competing for the same real resources, would have on the performance of these classes of applications combined with different types of libraries, aiming to determine which combinations could be consolidated into the same real resource and which combinations must be avoided. For the evaluation of the results, two types of analysis were conducted to examine the average performance loss (percent loss) and the distance between these losses (stability of the environment).

## 2  Literature Review

With the goal of categorizing the styles of computing used in scientific computing, the work of Colella [6] has identified seven numerical methods that he believed were important to science and engineering. Colella has introduced the "Seven Dwarfs" of scientific computing. Kaltofen [7] defines Dwarf as follows: "A dwarf is an algorithmic method that captures a pattern of computation and communication". This pattern is important in the proposed work to provide a basis of knowledge and characterization of types of applications so that when we assess that a certain application can be run in shared environments alongside other applications without degrading the shared resource, one deduces that other applications of the same type (dwarfs) may also compete for the same feature without degrading the resource. The Berkeley team in parallel computation extended these classifications to thirteen Dwarfs after they examined important application domains. They were interested in applying Dwarfs to a broader number of computational methods and investigating how well the Dwarfs could capture computation and communication patterns for a large range of applications [8]. The Dwarf classes being used in this paper are: Dense Linear Algebra (DLA), Structured Grid (SG) and Graph Transversal (GT) [11]. These three classes were chosen because there are a great number of scientific applications in various scientific fields, as shown in Figure 1.

|  | Embed | SPEC | DB | Games | ML | HPC | Health | Image | Speech | Music | Delaunay mesh generation | Gene Sequencing | Kmeans Clustering | Vacation Reservation System |
|---|---|---|---|---|---|---|---|---|---|---|---|---|---|---|
| Finite State Mach. | ■ | ■ | ■ | ■ | | | | | | | | | | |
| Combinational | ■ | | ■ | ■ | | | | | | | | ■ | | |
| **Graph Traversal** | ■ | ■ | ■ | ■ | | | ■ | ■ | ■ | | ■ | ■ | | ■ |
| **Structured Grid** | ■ | ■ | | ■ | | ■ | | ■ | | | | | | |
| **Dense Matrix** | ■ | ■ | ■ | | ■ | ■ | | | | | | | ■ | |
| Sparse Matrix | | | | ■ | ■ | | | | | ■ | | | | |
| Spectral (FFT) | ■ | | | | ■ | ■ | | | | | | | | |
| Dynamic Prog | ■ | | ■ | | | | | | | | | | | |
| N-Body | | | | | ■ | ■ | | | | | | | | |
| MapReduce | ■ | | | | ■ | | ■ | ■ | | | | | ■ | |
| Backtrack/ B&B | | | | | ■ | | | | | | | | | |
| Graphical Models | | | | | | | | | ■ | | | | | |
| Unstructured Grid | | | | ■ | | | ■ | | ■ | | | | | |

Figure 1: Examples of scientific areas characterized by Dwarf (http://stamp.stanford.edu). Relevance of class to domain applications. Emphasis on classes used in this work.

The benchmark used for the tests was Rodinia [9], a suite for heterogeneous computing which helps in the study of hardware platforms, particularly GPUs (Graphics Processing Units) and CPUs (Central Processing Units). The applications in the package are based on the Dwarf classifications [10]. This work tested OpenCL, or OCL, a parallel programming library that allows us to obtain a standard for writing applications that access all available programming resources, both on CPUs and GPUs as well as for other processors. OpenMP, or OMP, is an application programming interface (API) that is supported by most operating systems and processor architectures including Solaris, AIX, HP-UX, Linux, Mac OS X platforms and Windows and is based on a parallel programming model for shared memory multi-processor architectures [14].

Tanenbaum [12] defines process affinity relative to a CPU when there is a "smart algorithm" scaled to a CPU that has just ended another process. The time spent to exchange cache pages is reduced, justifying the affinity between that process and the CPU. However, this is not the same situation that happens in a cloud computing environment in which the processes are running concurrently on the same host sharing the same CPU. Some studies show affinity as a methodology to gain processing speed. These definitions also focus exclusively on real machines and only on the affinity between CPUs and memories rather than on the context of virtual machines in computational clouds. These details are shown in [13].

The effects of concurrency and degradation caused by processing in a virtualized environment that consumes all the CPU in the real server are shown in [2], where the effects on another process in a different virtualized environment are shown when the processes are allocated on the same real server and have to compete for the CPU, damaging processing capacity and performance.

With respect to the use of cloud computing, Calheiros et al. [3] present the CloudSim in their work, aiming to supply a system of general and extensible simulation, enabling modeling and simulation and providing a testing infrastructure for cloud computing and application services. According to the authors, using CloudSim, researchers and developers could focus on researching specific issues of system design without worrying about the low-level details, related infrastructure and services, but there is no assurance that the actual environment will follow the same forecast.

In this paper, the affinity concept is described as a role model in which scientific applications can run concurrently in virtualized environments on the same physical resource with minimum loss or with the ability to evaluate the loss. Research to date has shown that there is still a gap between evaluating the effects of concurrency in a virtualized environment and determining how to minimize these effects. Based on the Dwarf classification, the present work evaluates the effects of the concurrency among the classes of Dwarfs and the effects caused by the use of different types of parallel libraries, identifying those classes hosted in virtualized environments that could share the same real environment with minimal loss of performance. From a practical viewpoint, the results of this study could contribute to the development of schedulers able to use the concept of affinity to better distribute the requested virtual environments and therefore the improve the quality of services in consolidated environments or a cloud.

## 3 Methodology and Test Environment

To run the tests, three servers were created and configured with the following specifications: Intel(R) Xeon(R) CPU 2.67GHz X5650 - 12 cores, 24 Gb of RAM memory, Linux Ubuntu Server 12.04 operating system. On each real server, 3 virtual servers that are initialized simultaneously in the tests were created. These servers were configured as follows: QEMU Virtual CPU version 1.0 (Intel(R) Xeon(R) CPU 2.67GHz X5650 - 12 cores), 4 Gb of RAM memory, Linux Ubuntu Server 12.04 operating system. All servers shared processing resources in their 12 cores to verify concurrency applications.

The tests were performed with the following combinations of environments: i) Real Environment with OpenCL; ii) Virtual Environment with OpenCL; iii) Real Environment with OpenMP; iv) Virtual Environment with OpenMP; v) Real Environment with OpenCL X Virtual Environment with OpenMP; vi) Virtual Environment with OpenCL X Real Environment with OpenCL; vii) Real Environment with OpenCL X Real Environment with OpenMP; and viii) Virtual Environment with OpenCL X Virtual Environment with OpenMP. For each combination of the above, 20 samples were collected to analyze the performance in the tests, which proved sufficient to generate a reliable database. This number was achieved after assessing the confidence interval of the samples. In this case, we decided that, excluding the samples with constant times, the variations of the same runtime comparisons were not significant and remained in a close range with 10 runs. Therefore, to avoid unreliable conclusions, we chose to test each combination of algorithms, even those with constant runtimes, 20 times.

The tests were performed with the 4 types of algorithm (B+Tree, kmeans, LUD and SRAD) with combinations of two types of library (OpenMP and OpenCL) in real and virtual environments. Initially, the algorithms were tested in an environment free of concurrency to verify the runtimes, thus providing a baseline with which to compare the loss caused by the concurrency. The baseline was then compared with the concurrency of all 4 algorithms implemented using the OpenMP library in real and virtual environments and with all 4 algorithms implemented in OpenCL in real and virtual environments. All possible combinations were tested, disregarding the initial tests of each implementation, to adjust the tests (7120 tests), which generated a base of comprehensive and reliable knowledge.

## 4 Heterogeneous performance evaluation assessment

In this section, we will show the impact of concurrency among three classes of applications using four algorithms and compare the effects of the use of the OpenMP and OpenCL libraries. The results take into account the concurrence between two algorithms when using the two different libraries (cross-comparison). These cross library tests were classified in the affinity study as Heterogeneous Performance Evaluation Assessment.

The tests were conducted to measure the average performance loss and the impact on the stability of the environment caused by the concurrence to present the best and worst combinations among the algorithms when using OpenCL and OpenMP libraries. The distance between the performance averages was used to evaluate the stability of the environment, where a smaller distance represents a more stable environment and a greater distance represents a lower environmental stability. While the average performance loss will represent a quantitative criterion, the distance between the performance averages will represent a qualitative criterion for the environment and the tested algorithms.

### 4.1 LUD - Heterogeneous performance evaluation

Figure 2 shows the results with the LUD algorithm, as a baseline, running in a real or virtual environment under concurrency with other virtual or real environments hosting LUD. In the figure, following the order of appearance of the type of environment (R - Real, V-Virtual) and the type of algorithm, the order of the implemented libraries is always such that the first algorithm uses OpenCL and the second algorithm uses OpenMP. In the case of R.LUD x V.LUD, the real environment is running the LUD algorithm implemented with OpenCL and the virtual environment is running LUD implemented with OpenMP. This convention is used in all the subsequent evaluations.

In the figure, when the LUD algorithm is implemented with OpenCL libraries running in a virtual environment, the LUD algorithm had the least performance loss in all the concurrent scenarios in comparison with its implementation with OpenMP. A better quantitative combination was achieved when the LUD algorithm was implemented with OpenCL running in a real environment (105% loss) concurrently with another LUD algorithm implemented with OpenMP and also running in a real environment (37% loss), resulting in an average performance loss of 71%. This combination also had the least distance between the results (68%), representing the more stable combination.

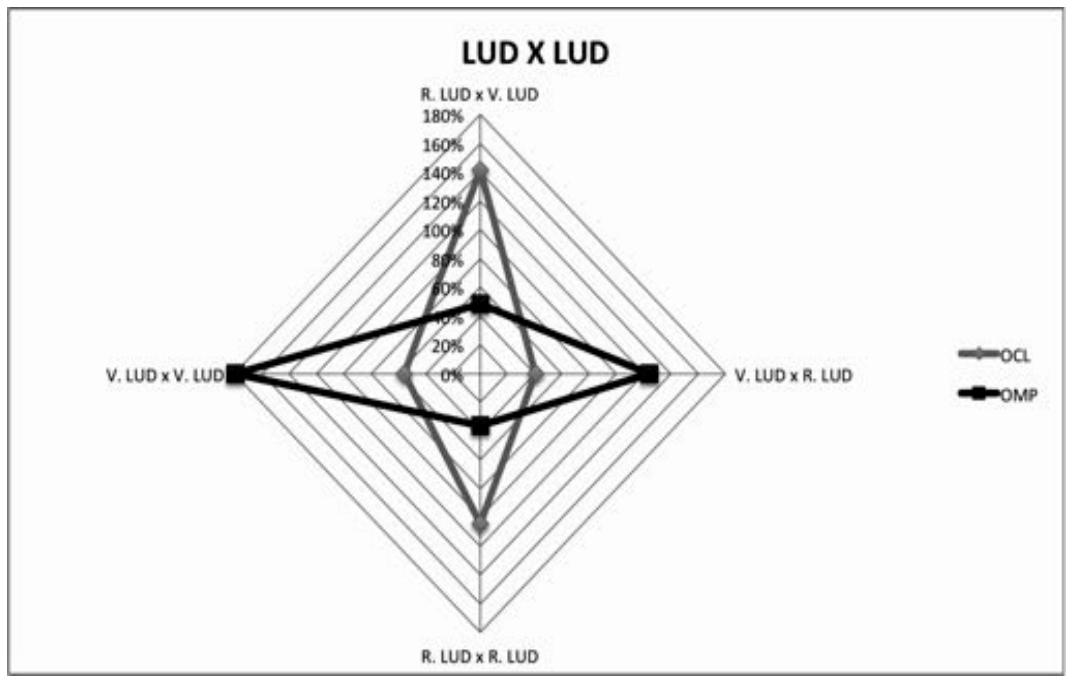

Figure 2: The performance loss in a real or virtual environment, caused by the concurrency between the LUD algorithm implemented with OpenCL and OpenMP libraries.

The worst combination was the virtual LUD implemented with OpenCL (55% loss) running concurrently with the virtual LUD implemented with OpenMP (179% loss). The average performance loss was 117% and the distance between the results was 124%.

## 4.2 B+Tree performance evaluation

Figure 3 shows the results with the B+Tree algorithm, as a baseline, running in a real or virtual environment concurrently with other virtual or real environments hosting B+Tree.

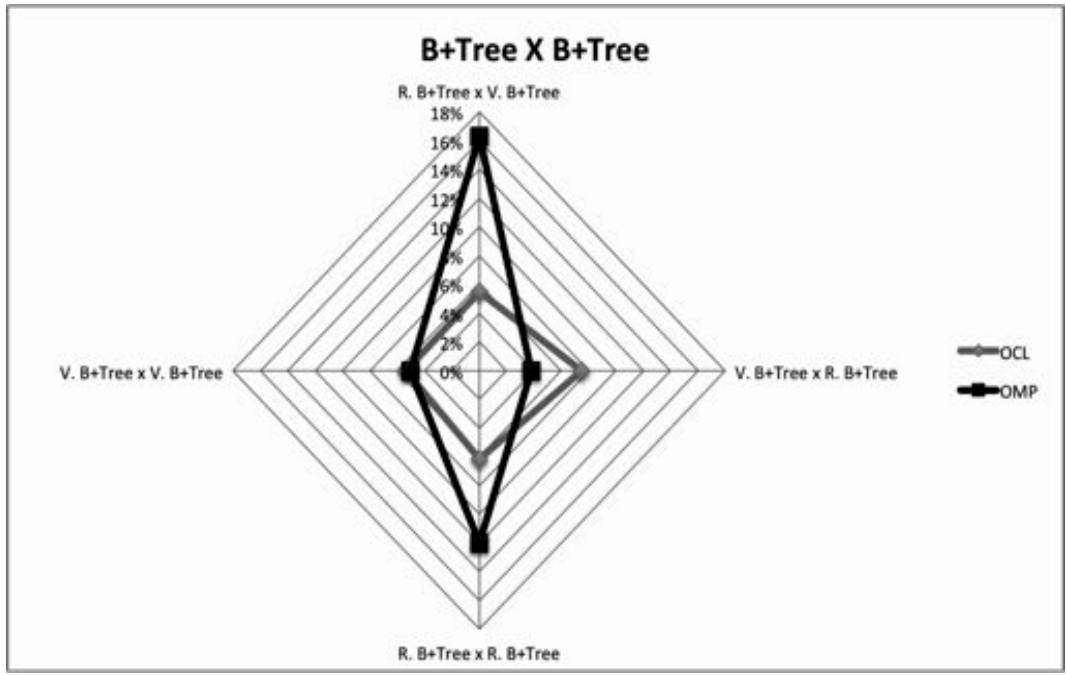

Figure 3: The performance loss in a real or virtual environment, caused by the concurrency between the B+Tree algorithm implemented with OpenCL and OpenMP libraries.

In the figure, the best performance combination was with B+Tree implemented with OpenCL and Open MP, both in a virtual environment with an average performance loss of 5% and the distance of 0 (zero) between them, representing the most stable combination.

The worst combination was real B+Tree implemented with OpenCL (5% loss) running concurrently with virtual B+Tree implemented with OpenMP (16% loss). The average performance loss was 11%, and the distance between the results was also 11%.

The proximity between the results shows the small influence of the libraries (OpenCL, OpenMP) in the performance loss of the B+Tree algorithm in the concurrent environment.

## 4.3 Kmeans performance evaluation

Figure 4 shows the results with the Kmeans algorithm, as a baseline, running in a real or virtual environment concurrently with other virtual or real environments hosting Kmeans.

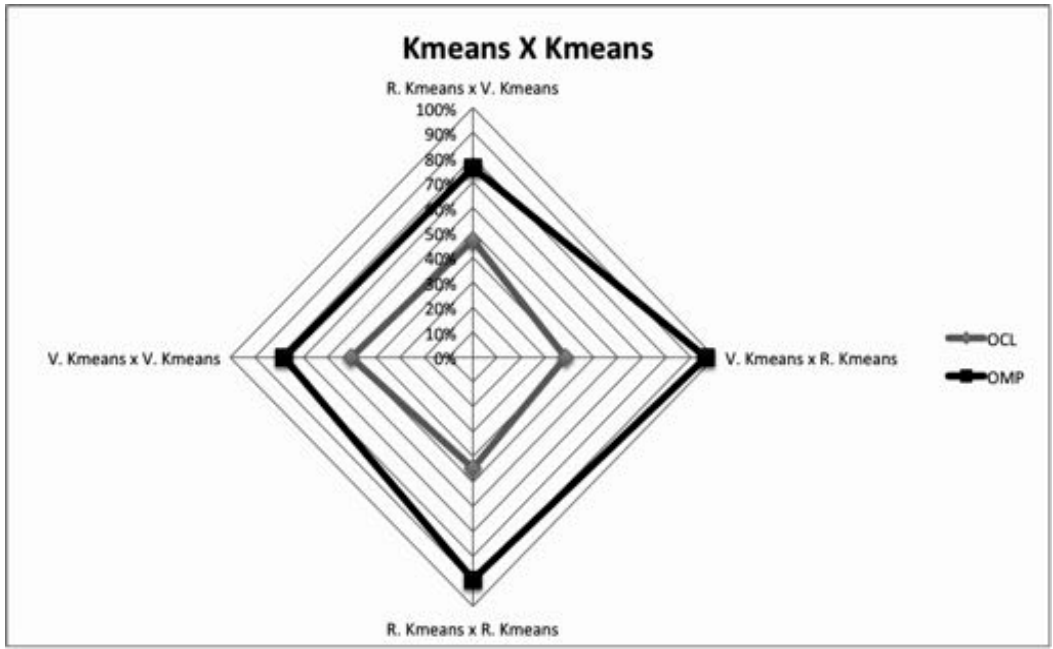

Figure 4: The performance loss in a real or virtual environment, caused by the concurrency between the Kmeans algorithm implemented with OpenCL and OpenMP libraries.

The figure shows that the least performance loss was achieved with the algorithm Kmeans implemented with OpenCL compared with Kmeans implemented with OpenMP for all environment combinations. The best combination was achieved with Kmeans implemented with OpenCL in a real environment (47% loss) running concurrently with Kmeans in a virtual environment implemented with OpenMP (76% loss), with an average performance loss of 61%. However, the least distance between the results was found with Kmeans implemented with OpenCL (50% loss) and with OpenMP (78% loss), both running in a virtual environment, with a distance of 28%.

In this example, there was one case with the best performance and one case with the best stability. Looking at the distance between Kmeans implemented with OpenCL in a real environment (47% loss) running concurrently with Kmeans in a virtual environment implemented with OpenMP (76% loss), the loss was 29%. Looking at the average performance loss of Kmeans implemented with OpenCL (50% loss) and with OpenMP (78% loss), each one in a virtual environment, the average performance loss was 64%. Considering the distance criterion, both are very close, but with respect to the performance criterion (based on the confidence interval), the best combination was achieved with Kmeans implemented with OpenCL in a real environment concurrently with Kmeans running in a virtual environment implemented with OpenMP.

## 4.4 SRAD performance evaluation

Figure 5 shows the results with the SRAD algorithm, as a baseline, running in a real or virtual environment concurrently with other virtual or real environments hosting SRAD.

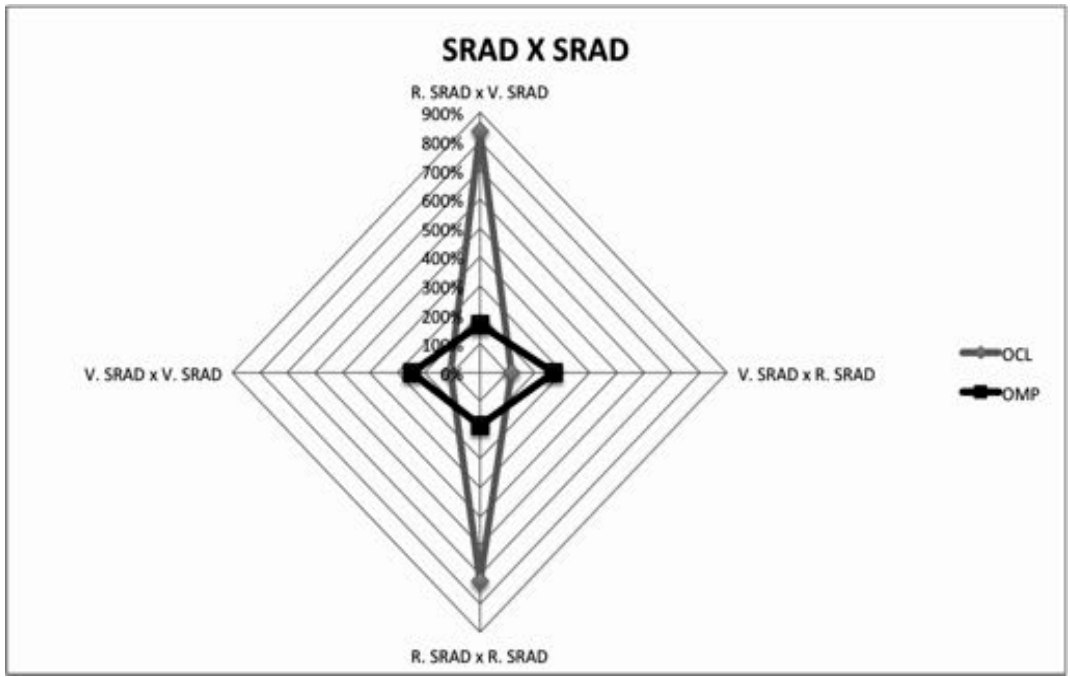

Figure 5: The performance loss in a real or virtual environment, caused by the concurrency between the SRAD algorithms implemented with OpenCL and OpenMP libraries.

The figure shows that the more stable environment was the SRAD algorithm implemented with OpenMP running concurrently with SRAD implemented with OpenCL, both in a virtual environment. However, none of the combinations of SRAD implemented with OpenCL running concurrently with another SRAD with OpenMP should be used, because there was an average performance loss between the algorithms of 176% for the best case (SRAD in a virtual environment implemented with OpenCL with 106% loss and OpenMP with 247% loss) with a distance of 141%.

The worst combination was the SRAD running in a real environment implemented with OpenCL (836% loss) running concurrently with SRAD in a virtual environment implemented with OpenMP (166% loss). The average performance loss was 501%, and this average performance loss represents the most unstable environment, with a distance of 669%.

## 4.5 Kmeans X LUD performance evaluation

Figure 6 shows the results with the Kmeans or LUD algorithm, as a baseline, running in a real or virtual environment concurrently with other virtual or real environments hosting Kmeans or LUD.

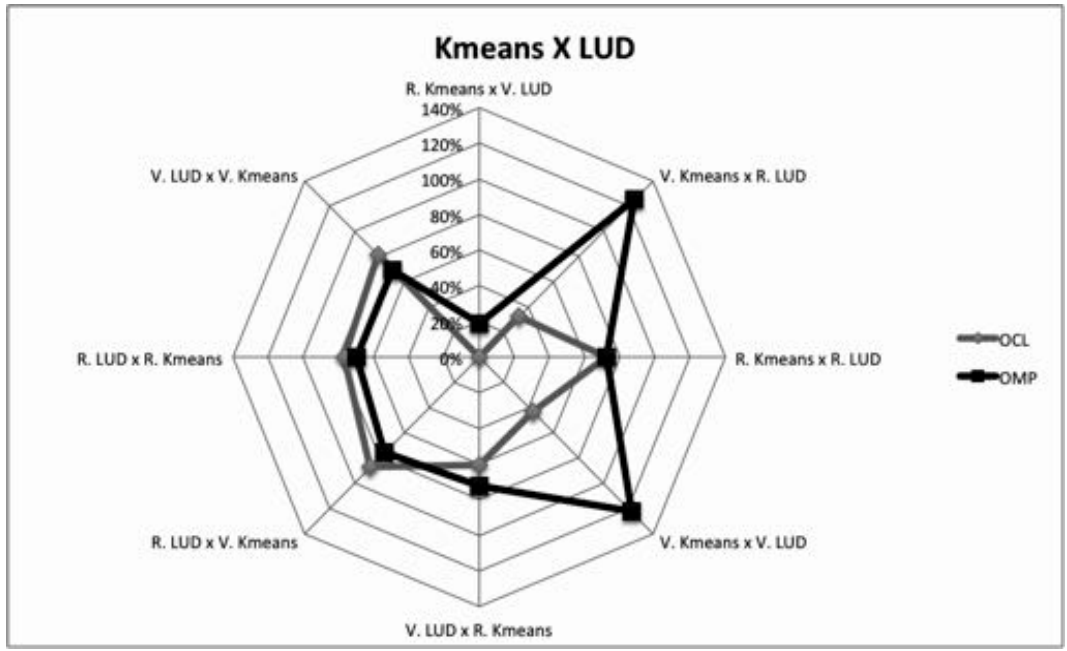

Figure 6: The performance loss in a real or virtual environment, caused by the concurrency between Kmeans or LUD algorithms implemented with OpenCL and OpenMP libraries

The figure shows that the best performance case was with Kmeans implemented with OpenCL executed in a real environment running concurrently with LUD implemented with OpenMP and executed in a virtual environment. In this case, the average performance loss was 9% (no loss for the Kmeans environment and 19% loss for the LUD environment).

However, this combination was not the most stable. The more stable combination was with Kmeans implemented with OpenCL (72% loss) running concurrently with LUD implemented with OpenMP (73% loss), with a distance of 1% between them, but in this case the average performance loss was 72%.

Based on the average performance loss, the best case was the first: Kmeans implemented with OpenCL executed in a real environment and running concurrently with LUD implemented with OpenMP executed in a virtual environment, with a distance of 19% between them.

The worst combination was Kmeans implemented with OpenCL executed in a virtual environment (43% loss) running concurrently with LUD implemented with OpenMP executed in a virtual environment (123% loss), with an average performance loss of 83%.

The most unstable combination was Kmeans implemented with OpenCL executed in a virtual environment (32% loss) running concurrently with LUD implemented with OpenMP executed in a real environment (125% loss), with a distance of 93%.

These results highlighted another interesting point related to the combination between Kmeans executed in a virtual environment concurrently with LUD executed in a real environment. Figure 6 shows that the combination of the LUD algorithm implemented with OpenCL libraries (87% loss) and Kmeans implemented with OpenMP libraries (76% loss) had a better stability, with a distance of 11% between them, compared to their reciprocal combination (LUD implemented with OpenMP libraries - 125% loss, and Kmeans implemented with OpenCL libraries - 32% loss), with a distance of 93%. These results are important because they validate the approach taken in this study, which considers the necessity of evaluating the effects of the composition between libraries, algorithms, and execution environments on performance and stability.

## 4.6 Kmeans X SRAD performance evaluation

Figure 7 shows the results with the Kmeans or SRAD algorithms, as a baseline, running in a real or virtual environment concurrently with other virtual or real environments hosting Kmeans or SRAD.

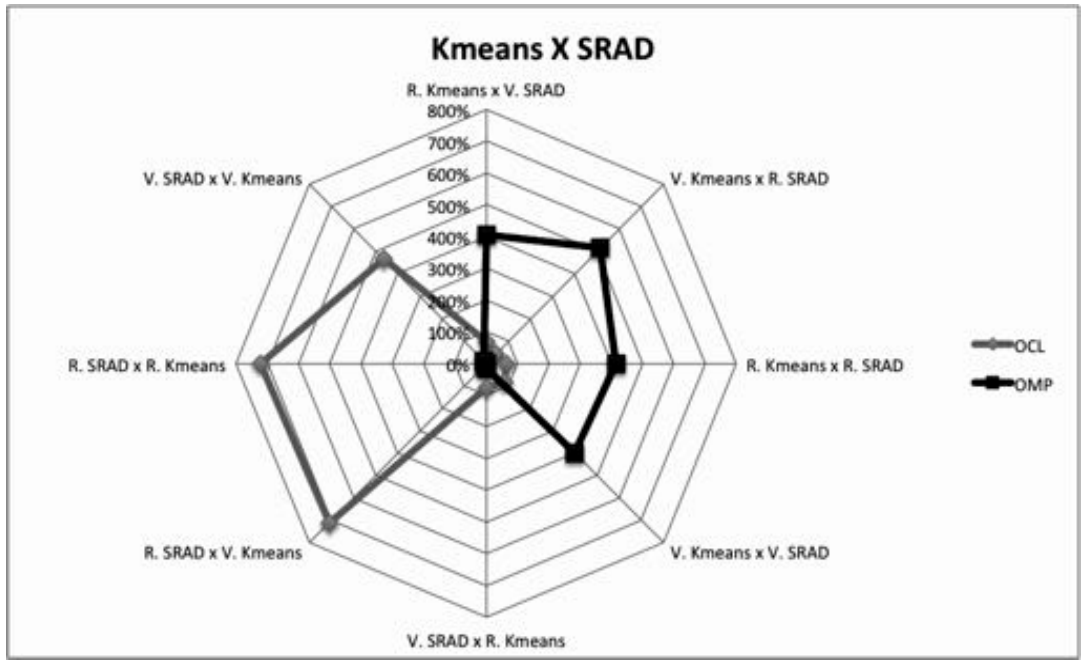

Figure 7: The performance loss in a real or virtual environment, caused by the concurrency between Kmeans or SRAD algorithms implemented with OpenCL and OpenMP libraries.

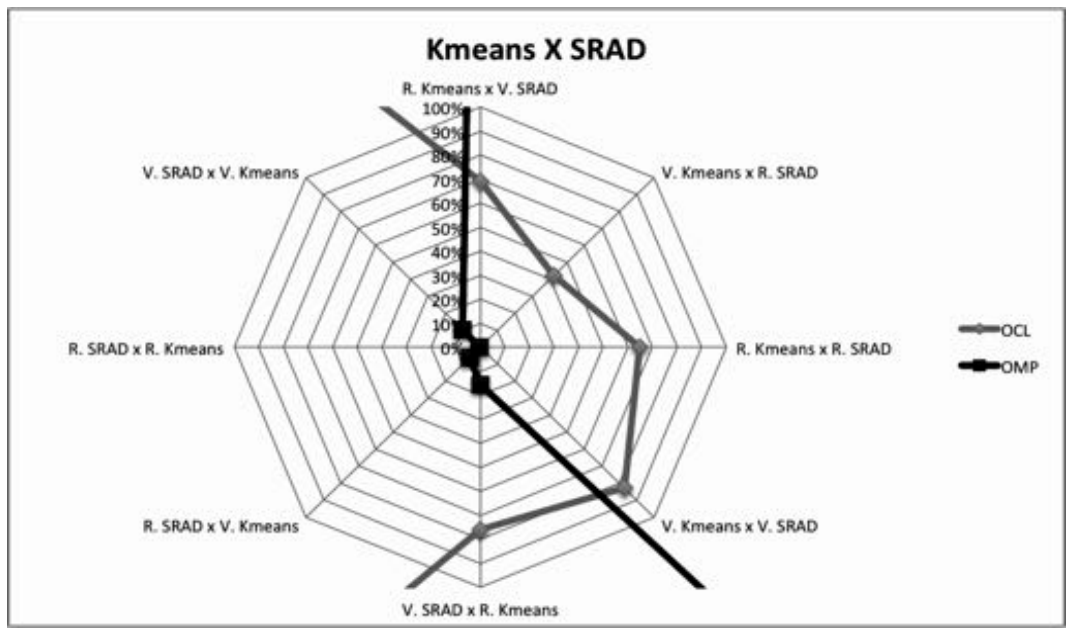

Figure 8: Detailed figure of the performance loss in a real or virtual environment, caused by the concurrency between Kmeans or SRAD algorithms implemented with OpenCL and OpenMP libraries.

Figure 8 shows, the best performance case was SRAD implemented with OpenCL executed in a virtual environment running concurrently with Kmeans implemented with OpenMP executed in a real environment. In this case, the average performance loss was 46% (76% loss for the SRAD environment and 15% loss for the Kmeans environment). This combination was also the most stable, with a distance between them of 61%.

The worst combination was SRAD implemented with OpenCL and executed in a real environment (721% loss) running concurrently with Kmeans implemented with OpenMP also executed in a real environment (0% loss), with an average performance loss of 361%. This was the most unstable environment, with a distance of 721%.

Figures 7 and 8 shows that the combination of the SRAD algorithm implemented with OpenCL libraries (76% loss) and Kmeans implemented with OpenMP libraries (15% loss) produced the best environment stability, with a distance of 61%. However, when SRAD was implemented with OpenMP libraries (404% loss) and Kmeans was implemented with OpenCL libraries (69% loss), the distance between increased to 335%, which was the worst stability, thus confirming the interaction and effects of the composition between libraries, algorithms and execution environments in determining the performance and stability.

## 4.7 Kmeans X B+Tree performance evaluation

Figure 9 shows the results with the Kmeans or B+Tree algorithms, as a baseline, running in a real or virtual environment with concurrency from other virtual or real environments hosting Kmeans or B+Tree.

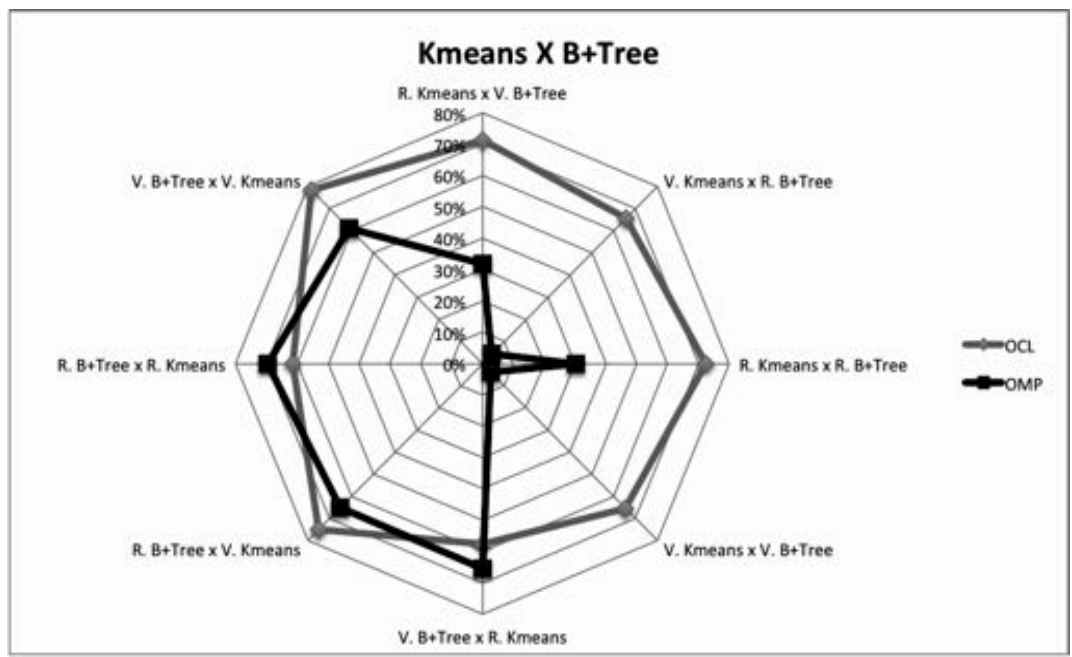

Figure 9: The performance loss in a real or virtual environment, caused by the concurrency between Kmeans or B+Tree algorithms implemented with OpenCL and OpenMP libraries.

The figure shows that the best performance case was Kmeans implemented with OpenCL executed in a virtual environment (65% loss) running concurrently with B+Tree implemented with OpenMP executed in a virtual environment (4% loss). In this case, the average performance loss was 35%. The most stable combination was B+Tree implemented with OpenCL executed in a virtual environment (58% loss) running concurrently with Kmeans implemented with OpenMP executed in a real environment (65% loss), with a distance between them of 7%.

In this case, there was a difficult trade-off between performance and stability. For the best performance, there was a larger distance (61%), and for the best stability there was an average performance loss of 60%. The evaluation and selection must be based on the optimization of the performance (first case) or on the optimization of the environment stability (second case).

The worst combination was B+Tree implemented with OpenCL executed in a virtual environment (78% loss) running concurrently with Kmeans implemented with OpenMP also executed in a virtual environment (61% loss), with an average performance loss of 70%.

The most unstable environment was Kmeans implemented with OpenCL executed in a virtual environment (66% loss) running concurrently with B+Tree implemented with OpenMP executed in a real environment (4% loss), with a distance between them of 62%.

When the Kmeans algorithm was running in a virtual environment and B+Tree was running in a real environment, if Kmeans was implemented with OpenMP (65% loss) and B+Tree was implemented with OpenCL libraries (75% loss), there was a better stability in the environment (distance of 10%) compared with its reciprocal (Kmeans implemented with OpenCL - 66% loss; B+Tree implemented with OpenMP - 4% loss), with a distance of 61%. All other combinations and their reciprocals showed the same behavior.

## 4.8 SRAD X LUD performance evaluation

Figure 10 shows the results with the SRAD or LUD algorithms, as a baseline, running in a real or virtual environment concurrently with other virtual or real environments hosting SRAD or LUD.

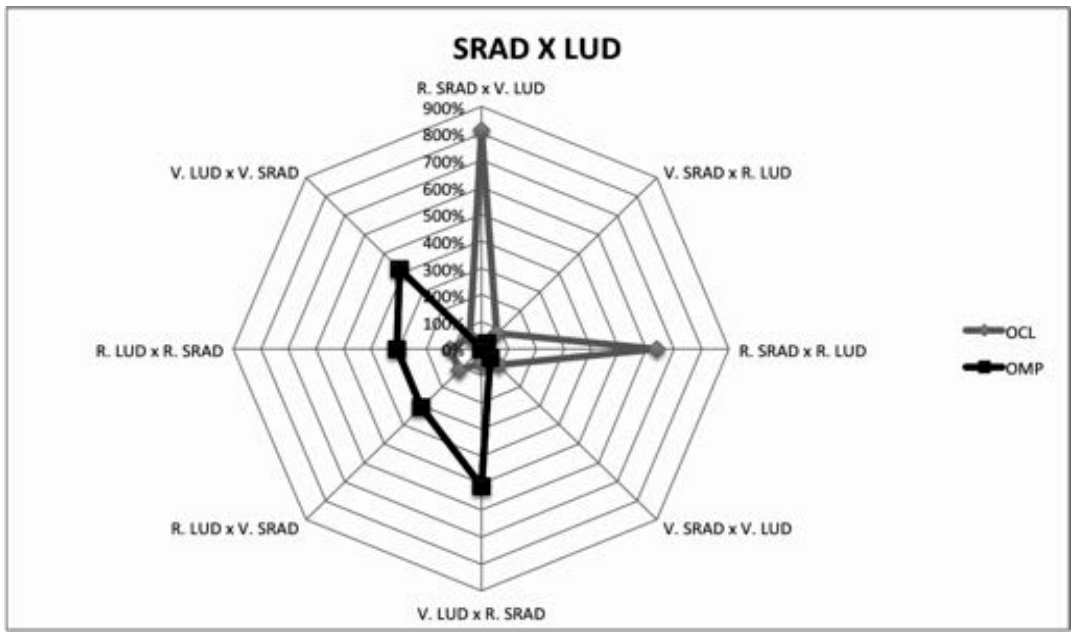

Figure 10: The performance loss in a real or virtual environment, caused by the concurrency between SRAD or LUD algorithms implemented with OpenCL and OpenMP libraries.

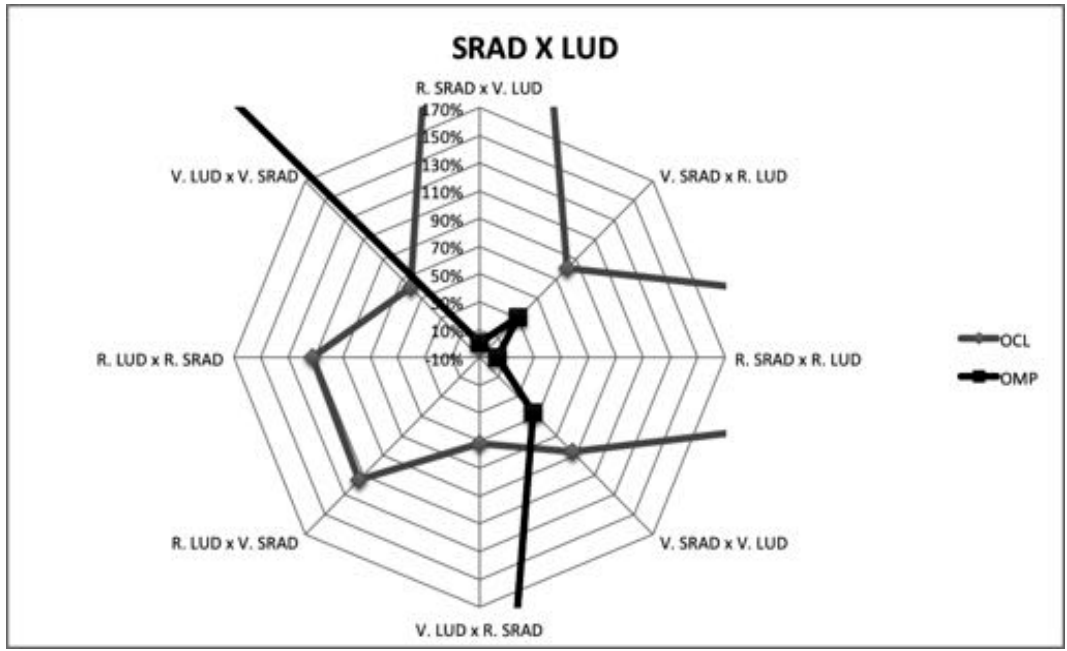

Figure 11: Detailed figure of the performance loss in a real or virtual environment, caused by the concurrency between SRAD or LUD algorithms implemented with OpenCL and OpenMP libraries.

Figure 11 shows that the best performance case was SRAD implemented with OpenCL executed in a virtual environment (81% loss) running concurrently with LUD implemented with OpenMP executed in a real environment (30% loss). In this case, the average performance loss was 56%.

The most stable combination was SRAD implemented with OpenCL executed in a virtual environment (87% loss) running concurrently with LUD implemented with OpenMP executed in a virtual environment (465% loss), with a distance between them of 41%.

In this case, there was also a difficult trade-off between performance and stability, because for the best performance (56% loss), there was the distance of 61%, while for the best stability (40%), there was an average performance loss of 66%. The evaluation and selection must therefore be based on the optimization of the performance (first case) or on the optimization of the environment stability (second case).

The worst combination was SRAD implemented with OpenCL executed in a real environment (811% loss) running concurrently with LUD implemented with OpenMP executed in a virtual environment (1% loss), with an average performance loss of 406%. This combination was also the most unstable combination of all the algorithms and libraries, with a distance of 810%.

## 4.9  B+Tree X LUD performance evaluation

Figure 12 shows the results with the B+Tree or LUD algorithms, as a baseline, running in a real or virtual environment concurrently with other virtual or real environments hosting B+Tree or LUD.

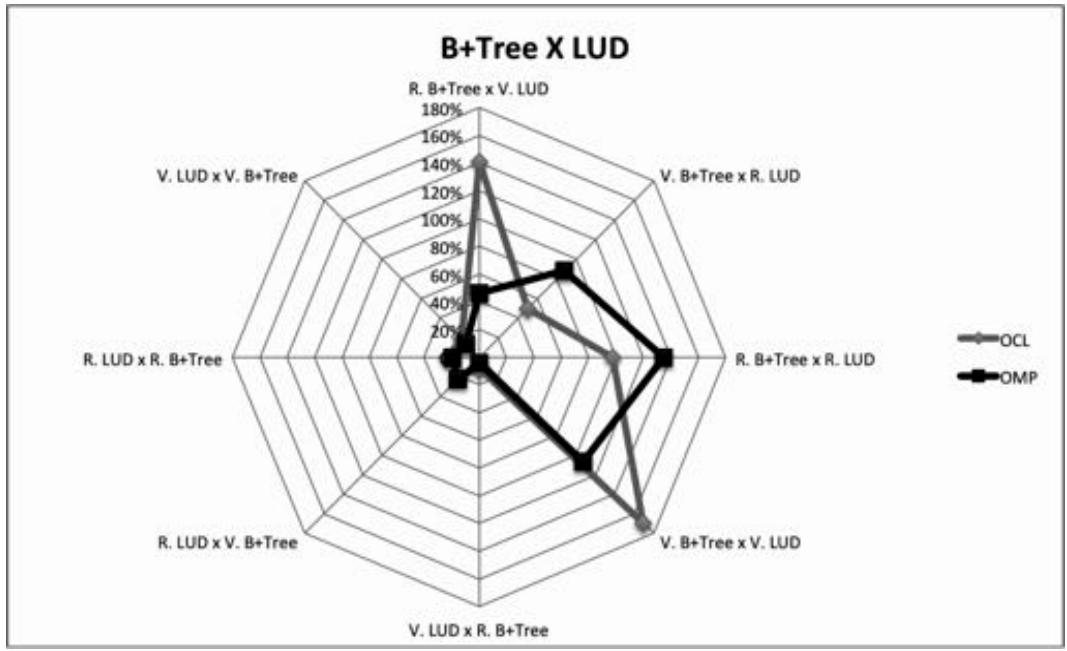

Figure 12: The performance loss in a real or virtual environment, caused by the concurrency between B+Tree or LUD algorithms implemented with OpenCL and OpenMP libraries.

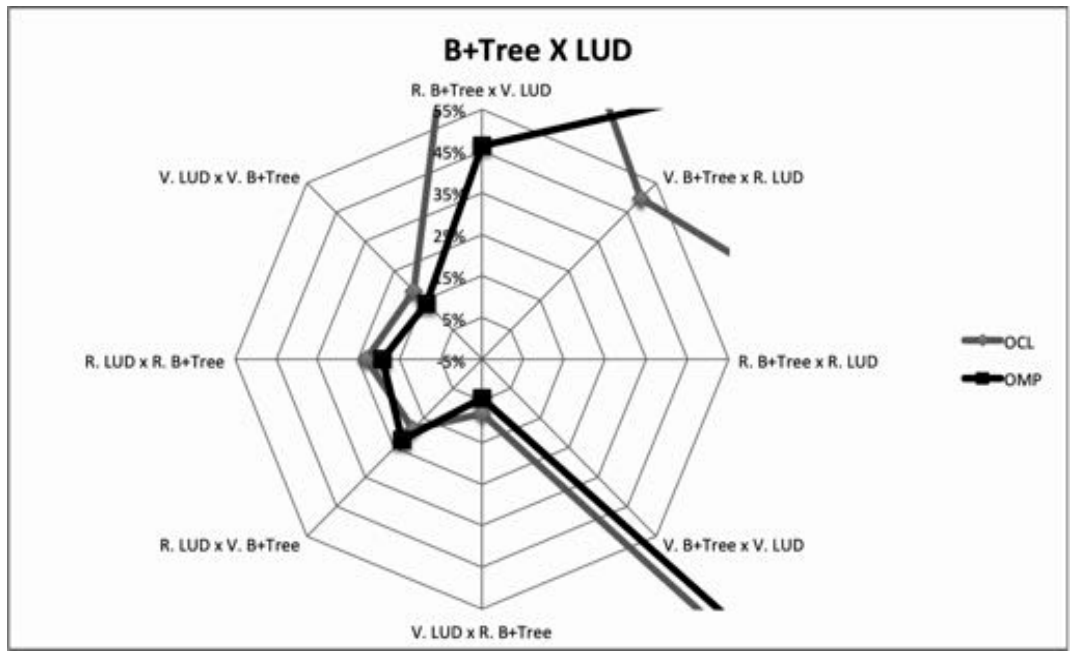

Figure 13: Detailed figure of the performance loss in a real or virtual environment, caused by the concurrency between B+Tree or LUD algorithms implemented with OpenCL and OpenMP libraries.

As Figure 13 shows, the best performance case was B+Tree implemented with OpenMP executed in a real environment (4% loss) running concurrently with LUD implemented with OpenCL executed in a virtual environment (8% loss). In this case, the average performance loss was 6%. This combination was also the most stable, with a distance of 4%.

The worst combination was B+Tree implemented with OpenCL executed in a virtual environment (170% loss) running concurrently with LUD implemented with OpenMP executed in a virtual environment (141% loss), with an average performance loss of 138%. The most unstable combination was B+Tree implemented with OpenCL executed in a real environment (141% loss) running concurrently with LUD implemented with OpenMP executed in a virtual environment (46% loss), with a distance of 95%.

The combination of these two algorithms and the libraries used in the tests becomes interesting because the LUD algorithm implemented with OpenCL libraries running concurrently in a real or virtual environment with B+Tree implemented with OpenMP libraries, also running in either environment, produces the least average performance loss and the best stability of all the combinations presented in this study. However, the reciprocal combination (LUD implemented with OpenMP and B+Tree implemented with OpenCL) produces results that are not as good.

## 4.10 B+Tree X SRAD performance evaluation

Figure 14 shows the results with the B+Tree or SRAD algorithms, as a baseline, running in a real or virtual environment concurrently with other virtual or real environments hosting B+Tree or SRAD.

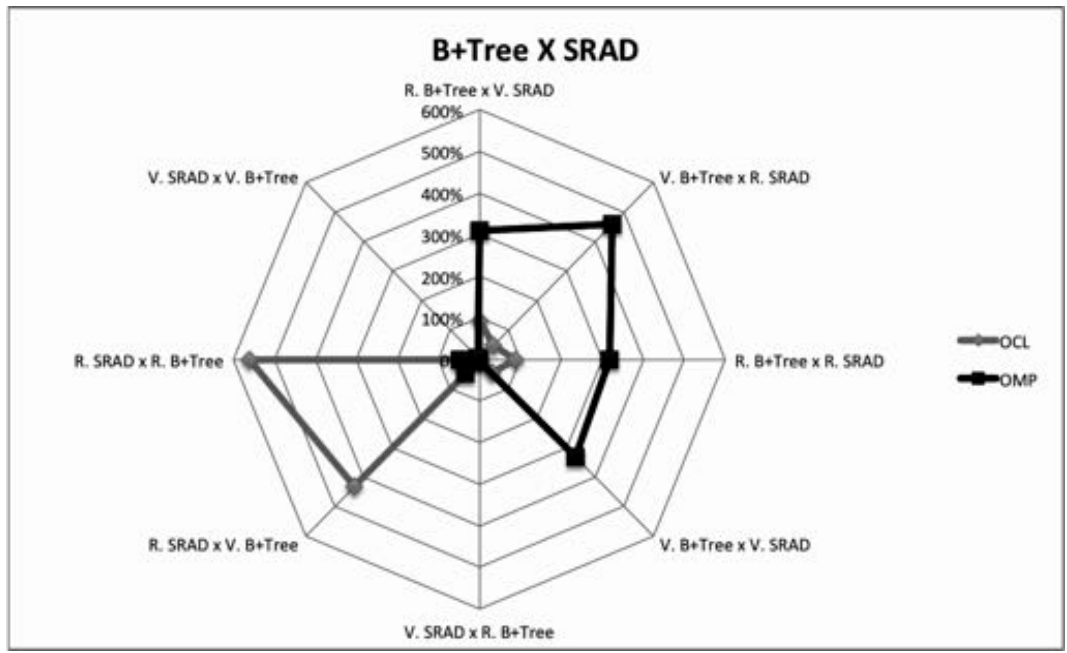

Figure 14: The performance loss in a real or virtual environment, caused by the concurrency between B+Tree or SRAD algorithms implemented with OpenCL and OpenMP libraries.

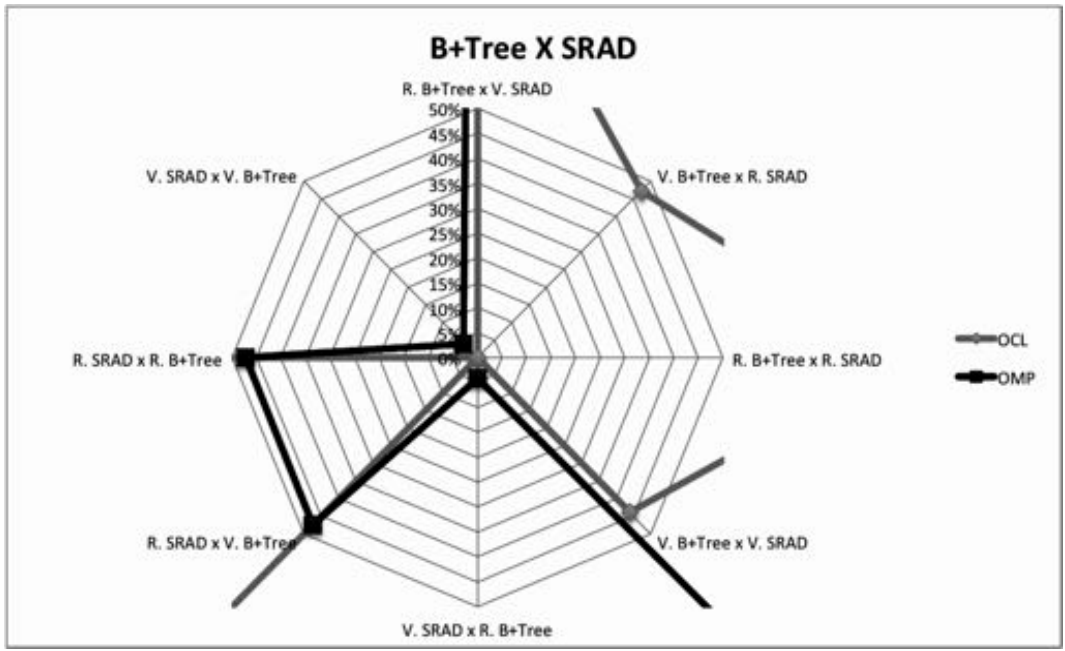

Figure 15: Detailed figure of the performance loss in a real or virtual environment, caused by the concurrency between B+Tree or SRAD algorithms implemented with OpenCL and OpenMP libraries.

As Figure 15 shows, there were two results with the least average performance loss. The first result was B+Tree implemented with OpenMP executed in a real environment (4% loss) running concurrently with SRAD implemented with OpenCL executed in a virtual environment (0% loss). The second result was B+Tree implemented with OpenMP executed in a virtual environment (0% loss) running concurrently with SRAD implemented with OpenCL executed in a virtual environment (4% loss). These were also the most stable combinations, with distances of 4%.

The worst combination was B+Tree implemented with OpenMP executed in a real environment (47% loss) running concurrently with SRAD implemented with OpenCL executed in a real environment (561% loss), with an average performance loss of 304%. This combination was also the most unstable, with a distance of 5%.

The combination of these two algorithms and the libraries used in the tests becomes interesting because the LUD algorithm implemented with OpenCL libraries running concurrently in a real or virtual environment with B+Tree implemented with OpenMP libraries, also running in either environment, produced the smallest average performance loss and the best stability of all the combinations presented in this study. However, the reciprocal combination (LUD implemented with OpenMP and B+Tree implemented with OpenCL) produced results that were not as good.

## 4.11 Library Results Evaluation

Figures 16, 17, 18 and 19 consolidate all the combinations and the results presented in the previous sections. The percentage in each bar represents the average performance loss due to concurrency considering three parameters: the heterogeneous library composition (OpenMP and OpenCL), the type of algorithm (LUD, SRAD, B+Tree and Kmeans) and the type of environment (real or virtual). The convention utilized in the figure is based on the sequence of the composition of the environments and the types of algorithm (example: V.LUD x R.SRAD for LUD running in a virtual environment concurrent with SRAD running in a real environment), with the first bar representing the loss suffered by the first composition implemented in OMP and the second bar representing the loss suffered by the second composition implemented in OpenCL.

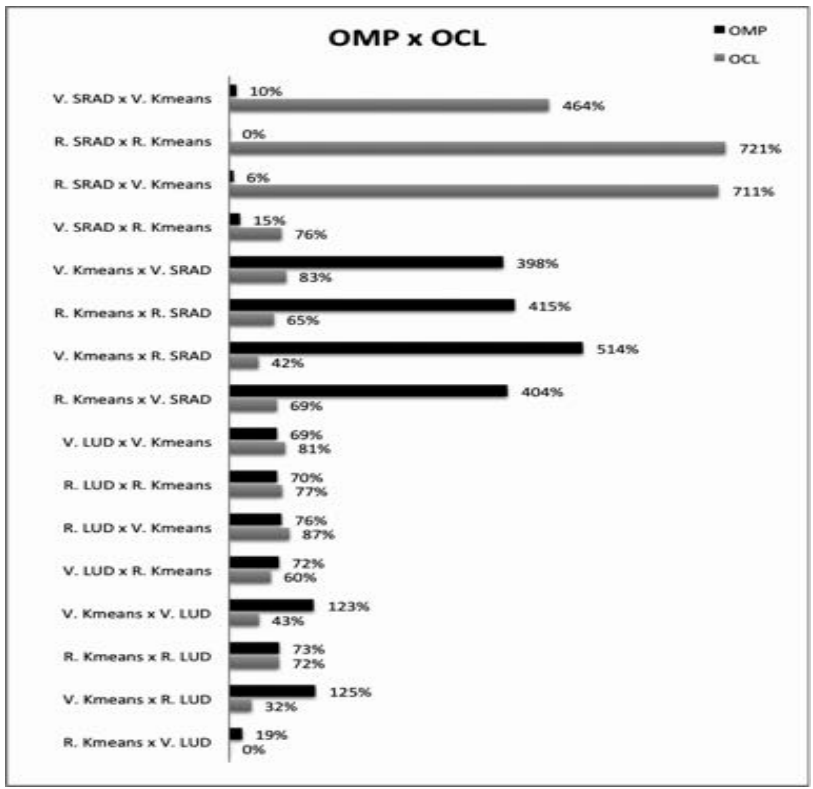

Figure 16: Impact results - OpenMP X OpenCL (Figure 1/4).

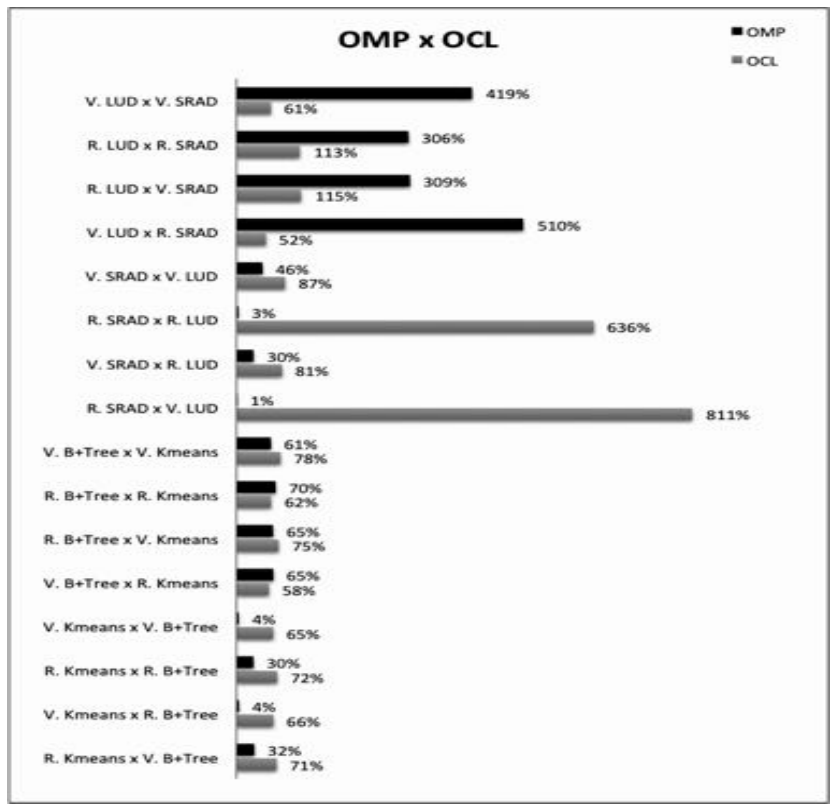

Figure 17: Impact results - OpenMP X OpenCL (Figure 2/4).

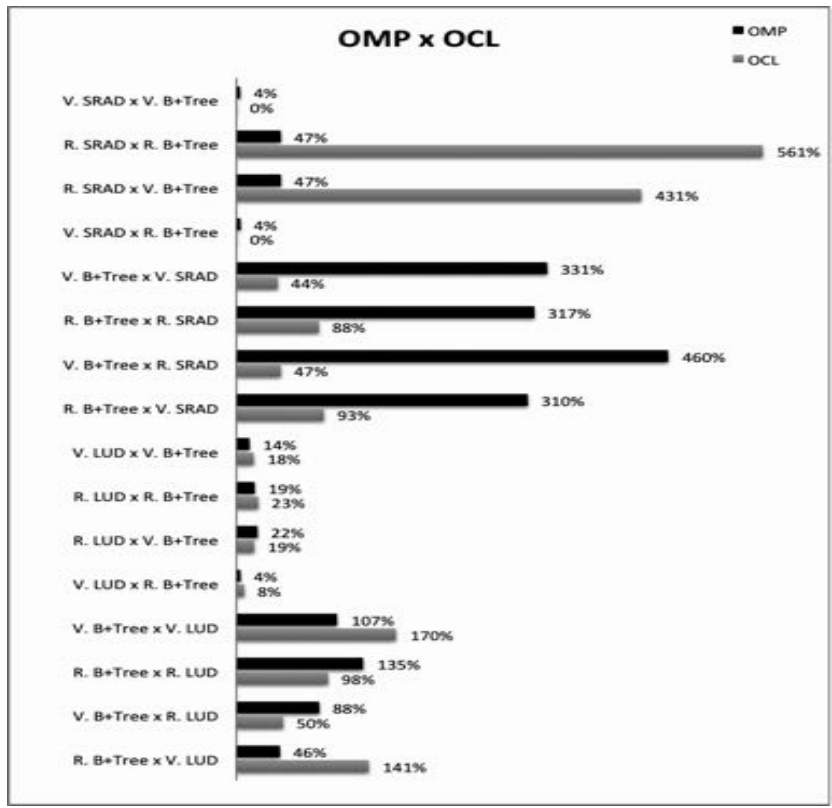

Figure 18: Impact results - OpenMP X OpenCL (Figure 3/4).

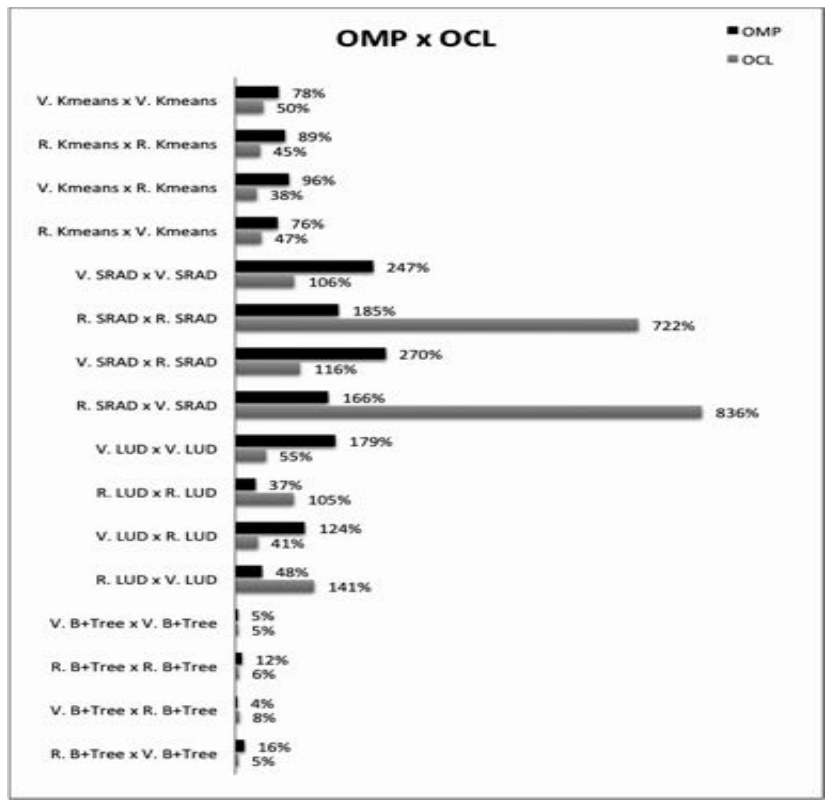

Figure 19: Impact results - OpenMP X OpenCL (Figure 4/4).

Figures 20, 21, 22, 23, 24, 25, 26 and 27 consolidate all impacts suffered for each of the algorithms with the same convention as stated before. In the next sections, a summary of the results will be presented for each of the algorithms. For comparison of these results, all the execution time values presented in the figures were normalized to allow the comparison of the distinct time values of each test. A scale ranging from 0 to 9 was used for the comparison.

## 4.12 Analysis of the impact suffered and caused by each algorithm running in a real environment due the effects of concurrency, and libraries

.
Figure 20 shows the impact caused or suffered by the concurrency of the Kmeans algorithm in a real environment with the other algorithms. The figure shows that the largest impact caused by Kmeans on the other algorithms occurred when it was running in a real environment implemented with OpenMP libraries, especially when this concurrency was with the B+Tree algorithm running in a real environment and implemented with OpenCL libraries. The normalized performance loss suffered by Kmeans was 3.30, and the normalized performance loss suffered by B+Tree was 7.65. In the case of Kmeans implemented with OpenCL, the largest impact was caused when running in a real environment concurrently with B+Tree implemented with OpenMP running in a virtual environment. The normalized performance loss of Kmeans was 5.85, and the normalized performance loss of B+Tree was 7.65. The greatest impact suffered by Kmeans caused by another algorithm occurred when Kmeans implemented with OpenCL libraries ran concurrently in a real environment with the LUD algorithm implemented with OpenMP and also running in a real environment. The normalized performance loss suffered by Kmeans was 7.09, and that suffered by LUD was 4.50.

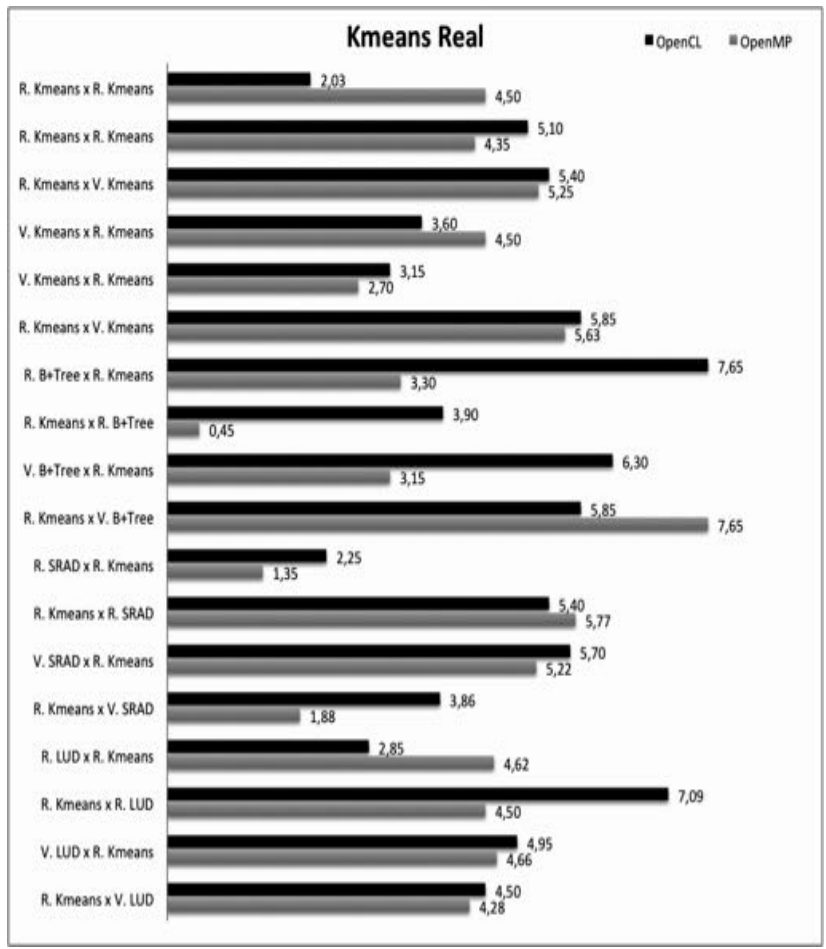

Figure 20: Impact caused or suffered by the concurrency of the Kmeans algorithm in a real environment

Figure 21 shows the impact caused or suffered by the concurrency of the Kmeans algorithm in a virtual environment with the other algorithms. The figure shows that the largest impact caused by Kmeans on the other algorithms occurred when it was running in a virtual environment implemented with OpenCL libraries, especially when this concurrency was with the SRAD algorithm running in a virtual environment and implemented with OpenMP libraries. The normalized performance loss suffered by Kmeans was 3.60, and that suffered by B+Tree was 8.10. The greatest impact suffered by Kmeans caused by another algorithm occurs when Kmeans implemented with OpenMP libraries ran concurrently in a virtual environment with the SRAD algorithm running in a real environment and implemented with OpenCL. The normalized performance loss suffered by Kmeans was 7.65, and the normalized performance loss suffered by LUD was 5.4.

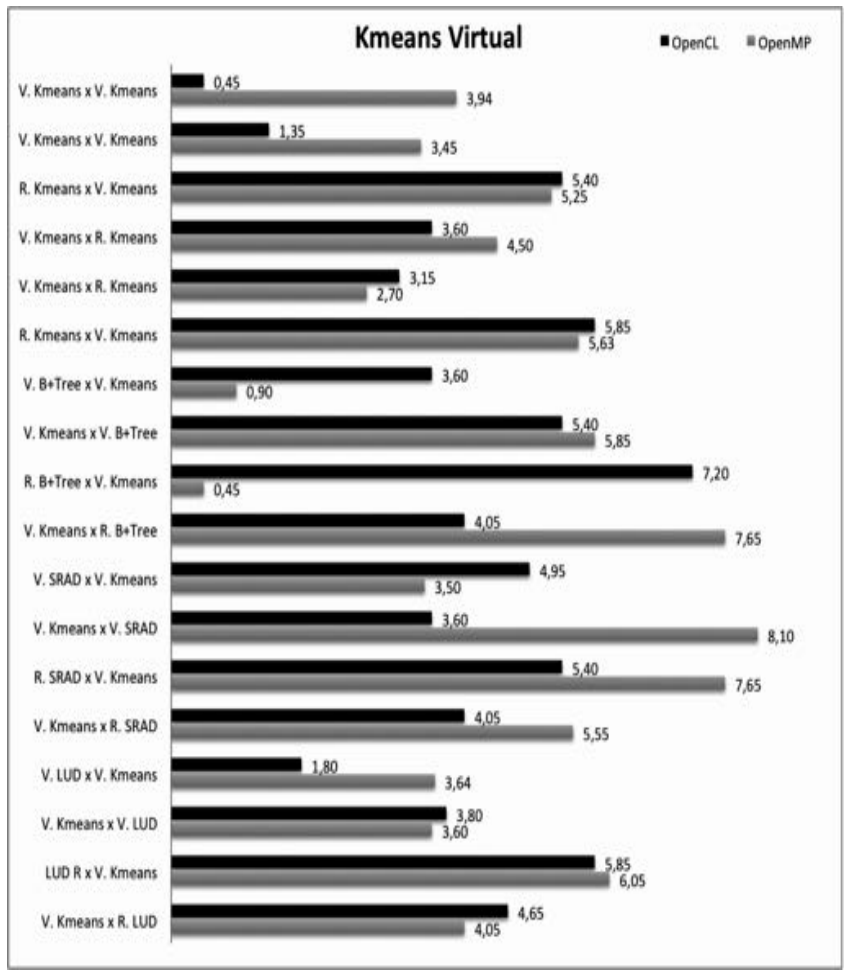

Figure 21: Impact caused or suffered by the concurrency of the Kmeans algorithm in a virtual environment

Figure 22 shows the impact caused or suffered by the concurrency of the LUD algorithm in a real environment with the other algorithms. The figure shows that the largest impact caused by LUD on the other algorithms occurred when LUD was running in a real environment implemented with OpenMP libraries, especially when this concurrency was with the Kmeans algorithm running in a real environment and implemented with OpenCL libraries. The normalized performance loss suffered by LUD was 4.50, and that suffered by Kmeans was 7.09. The greatest impact suffered by LUD caused by another algorithm occurred when LUD implemented with OpenCL libraries ran concurrently in a real environment with the LUD algorithm running in a real environment implemented with OpenMP. The normalized performance loss suffered by LUD with OpenCL was 6.75, and that suffered by LUD with OpenMP was 4.07.

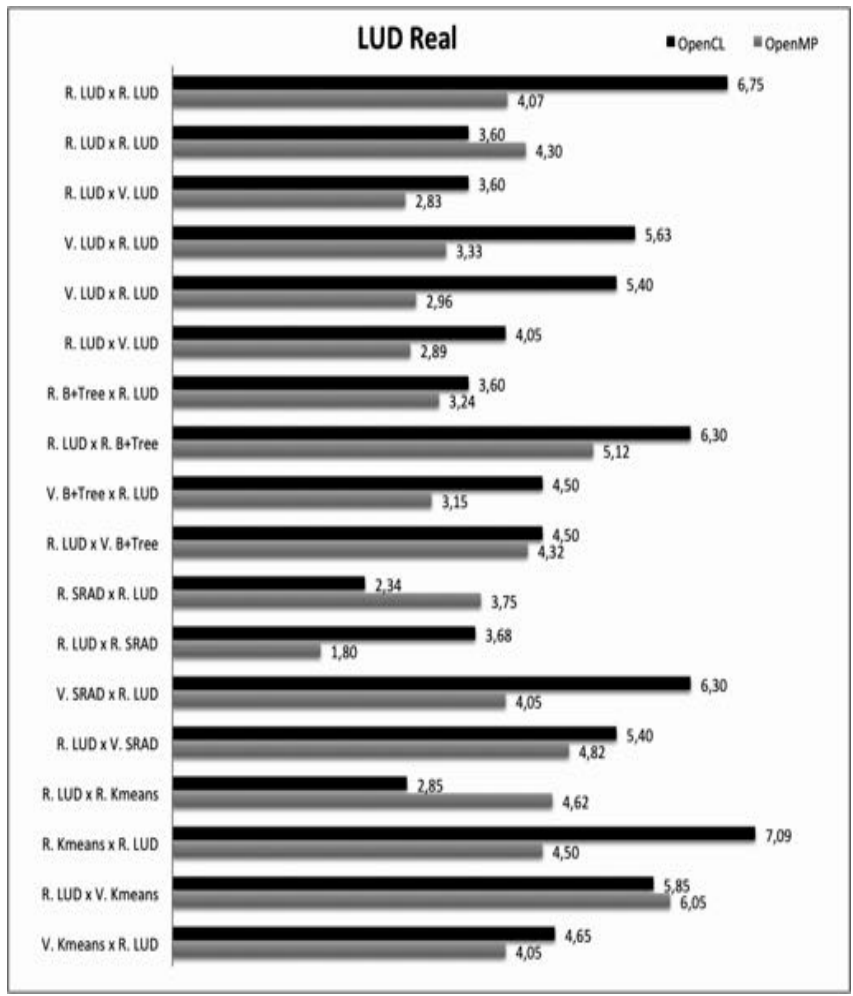

Figure 22: Impact caused or suffered by the concurrency of the Kmeans algorithm in a real environment

Figure 23 shows the impact caused or suffered by the concurrency of the LUD algorithm in a virtual environment with the other algorithms. The figure shows that the largest impact caused by LUD on the other algorithms occurred when it was running in a virtual environment implemented with OpenCL libraries, especially when this concurrency was with the B+Tree algorithm running in a virtual environment and implemented with OpenMP libraries. The normalized performance loss suffered by LUD was 4.32, and that suffered by B+Tree was 6.30. The greatest impact suffered by LUD caused by another algorithm occurred when LUD was implemented with OpenCL libraries and ran concurrently in a virtual environment with the LUD algorithm running in a real environment and implemented with OpenMP. The normalized performance loss suffered by LUD with OpenCL was 5.63, and that suffered by LUD with OpenMP was 3.33.

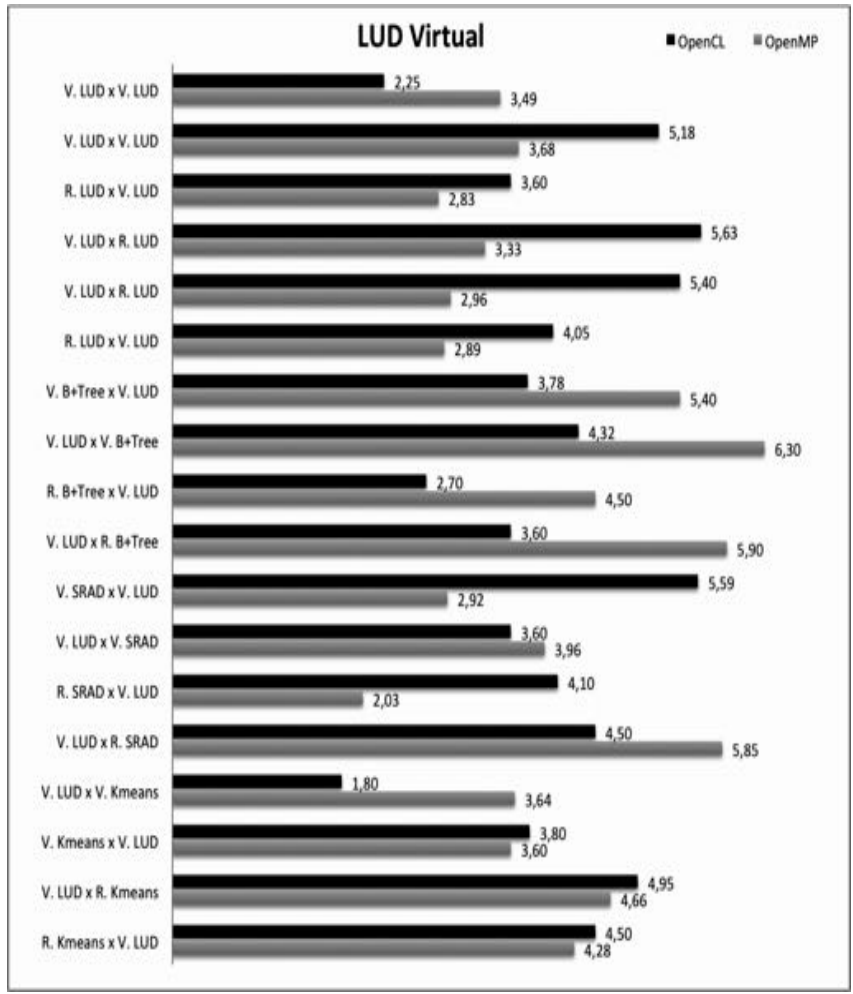

Figure 23: Impact caused or suffered by the concurrency of the LUD algorithm in a virtual environment

Figure 24 shows the impact caused or suffered by the concurrency of SRAD algorithm in a real environment with the other algorithms. The figure shows that the largest impact caused by SRAD on the other algorithms occurred when SRAD was running in a real environment implemented with OpenCL libraries, especially when this concurrency was with the Kmeans algorithm running in a virtual environment and implemented with OpenMP libraries. The normalized performance loss suffered by SRAD was 5.40, and that suffered by Kmeans was 7.65. The greatest impact suffered by SRAD caused by another algorithm occurred when SRAD was implemented with OpenMP libraries and ran concurrently in a real environment with the B+Tree algorithm running in a real environment implemented with OpenCL. The normalized performance loss suffered by SRAD was 8.55, and that suffered by B+Tree was 4.50.

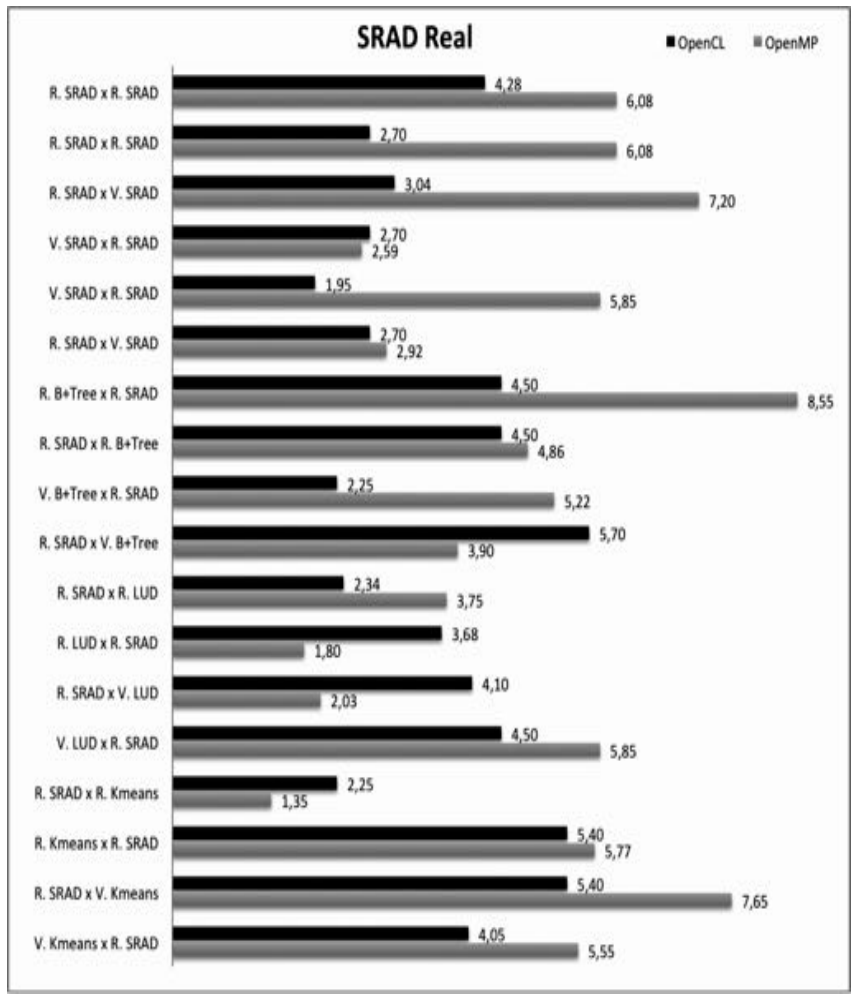

Figure 24: Impact caused or suffered by the concurrency of the SRAD algorithm in a real environment

Figure 25 shows the impact caused or suffered by the concurrency of the SRAD algorithm in a virtual environment with the other algorithms. The figure shows that the largest impact caused by SRAD on the other algorithms occurred when SRAD was running in a virtual environment implemented with OpenCL libraries, especially when this concurrency was with the SRAD algorithm running in a real environment and implemented with OpenMP libraries. The normalized performance loss suffered by SRAD with OpenCL was 1.95, and that suffered by SRAD with OpenMP was 5.85. The greatest impact suffered by SRAD caused by another algorithm occurred when SRAD implemented with OpenMP libraries ran concurrently in a virtual environment with the Kmeans algorithm running in a virtual environment implemented with OpenCL. The normalized performance loss suffered by SRAD was 8.10, and that suffered by Kmeans was 3.60.

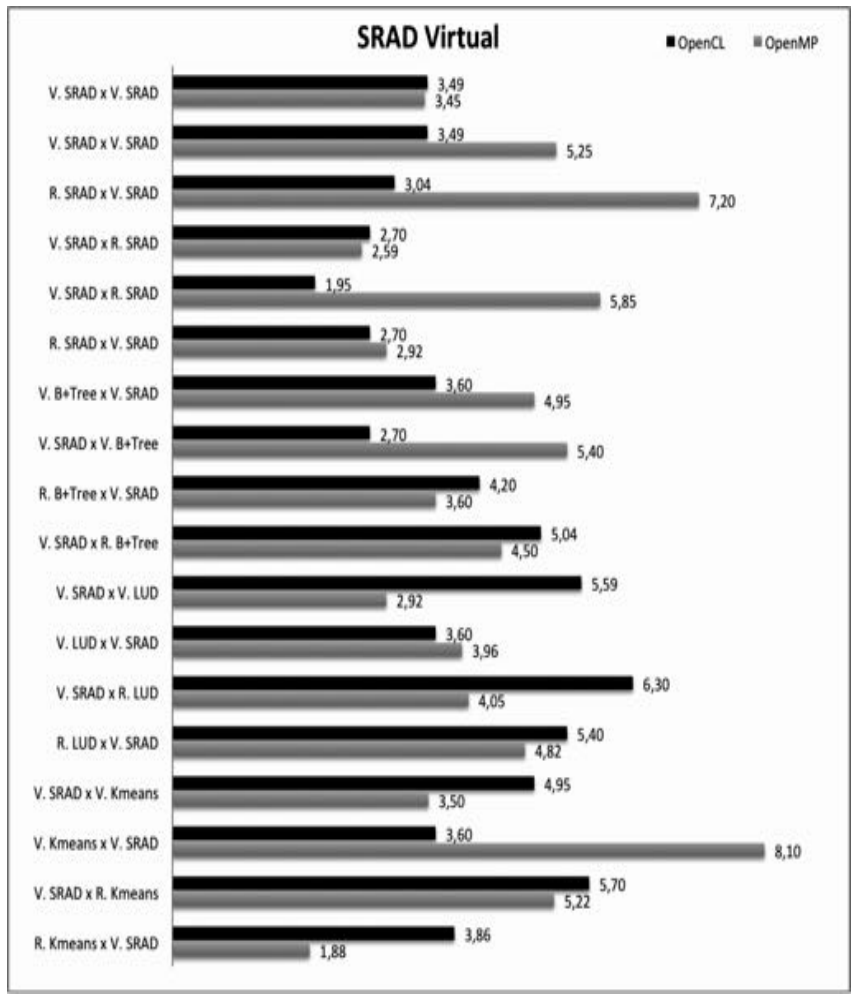

Figure 25: Impact caused or suffered by the concurrency of the SRAD algorithm in a virtual environment

Figure 26 shows the impact caused or suffered by the concurrency of the B+Tree algorithm in a real environment with the other algorithms. The figure shows that the largest impact caused by the B+Tree on the other algorithms occurred when B+Tree was running in a real environment implemented with OpenCL libraries, especially when this concurrency was with the SRAD algorithm running in a real environment and implemented with OpenMP libraries. The normalized performance loss suffered by B+Tree was 4.50, and that suffered by SRAD was 8.55. The greatest impact suffered by the B+Tree caused by another algorithm occurred when B+Tree implemented with OpenCL libraries ran concurrently in a real environment with the Kmeans algorithm running in a real environment implemented with OpenMP. The normalized performance loss suffered by B+Tree was 7.65, and that suffered by Kmeans was 3.30. Another combination with the same loss was B+Tree implemented with OpenMP libraries running concurrently in a real environment with the Kmeans algorithm running in a virtual environment implemented with OpenCL. The normalized performance loss suffered by B+Tree was 7.65, and that suffered by Kmeans was 4.05.

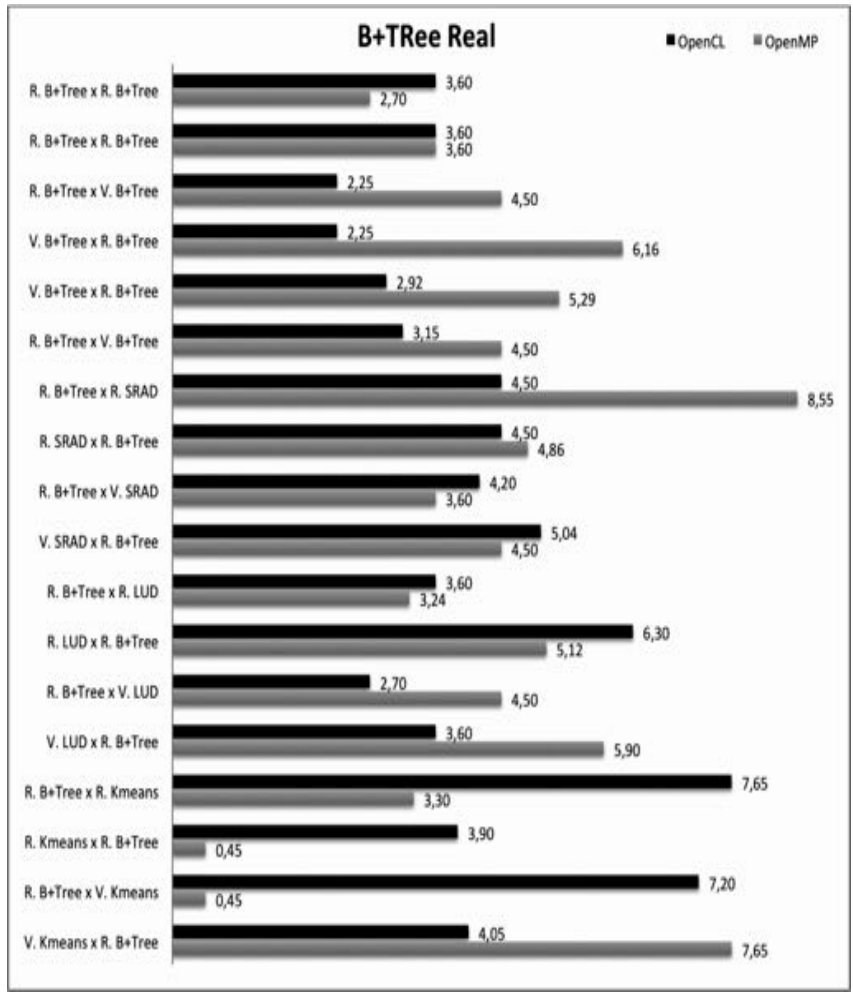

Figure 26: Impact caused or suffered by the concurrency of the B+Tree algorithm in a real environment

Figure 27 shows the impact caused or suffered by the concurrency of the B+Tree algorithm in a virtual environment with the other algorithms. The figure shows that the largest impact caused by B+Tree on the other algorithms occurred when B+Tree was running in a virtual environment implemented with OpenCL libraries, especially when this concurrency was with the B+Tree algorithm running in a real environment and implemented with OpenMP libraries. The normalized performance loss suffered by B+Tree with OpenCL was 2.50, and that suffered by B+Tree with OpenMP was 6.16. The greatest impact suffered by B+Tree caused by another algorithm occurred when B+Tree implemented with OpenMP libraries ran concurrently in a virtual environment with the Kmeans algorithm running in a real environment implemented with OpenCL. The normalized performance loss suffered by B+Tree was 7.65, and that suffered by Kmeans was 5.85.

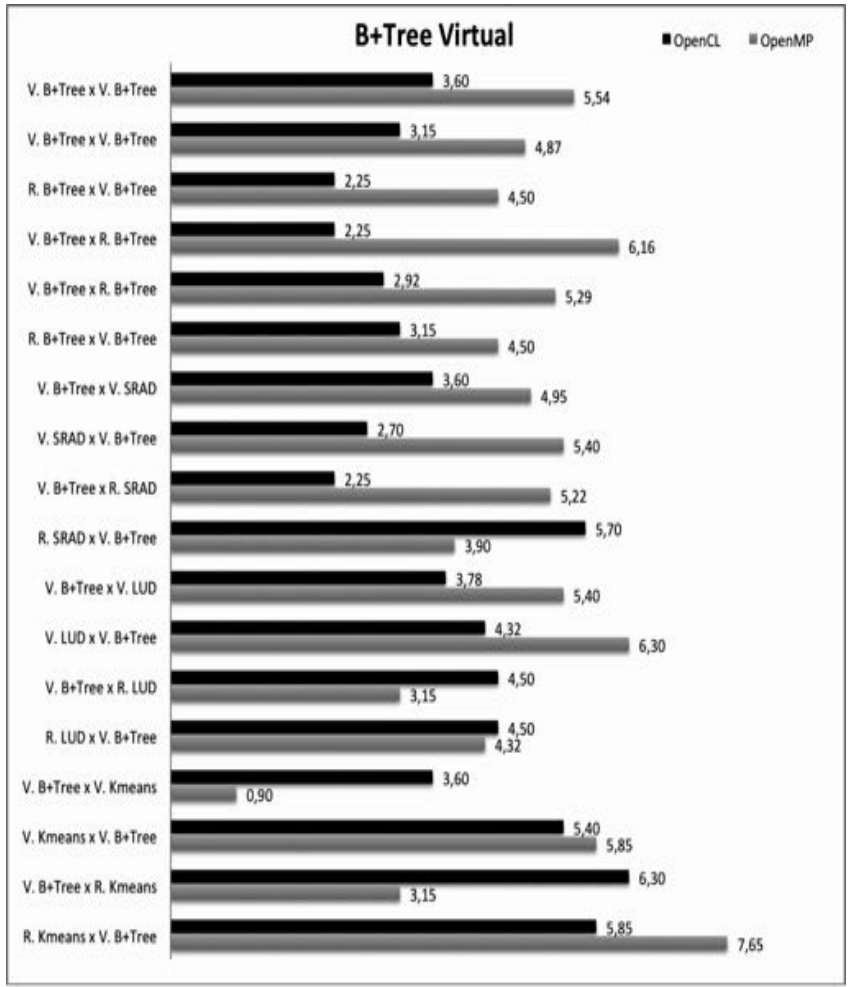

Figure 27: Impact caused or suffered by the concurrency of the B+Tree algorithm in a virtual environment

# 5  Consolidation of Results

A heterogeneous evaluation of the results presented in Figure 28 shows the comparative percentage of impacts suffered by each algorithm implemented with the OpenCL or OpenMP library. In the figure, it is apparent which algorithms can coexist in the same physical resource and which cannot. The representation used in this figure gives the percentage loss of the implemented algorithm with the OpenCL or OpenMP library, and a more darkly shaded cell indicates a worse loss associated with the combination in question.

| X | OpenMP | | | | | | | |
|---|---|---|---|---|---|---|---|---|
| | Real LUD | Virtual LUD | Real Kmeans | Virtual Kmeans | Real SRAD | Virtual SRAD | Real B+Tree | Virtual B+Tree |
| **OpenCL** Real LUD | 37% / 105% | 48% / 141% | 70% / 77% | 76% / 87% | 306% / 113% | 309% / 115% | 19% / 23% | 22% / 19% |
| Virtual LUD | 124% / 41% | 179% / 55% | 72% / 60% | 69% / 81% | 510% / 52% | 419% / 61% | 4% / 8% | 14% / 18% |
| Real Kmeans | 73% / 72% | 19% / 0% | 89% / 45% | 76% / 47% | 415% / 65% | 404% / 69% | 30% / 72% | 32% / 71% |
| Virtual Kmeans | 125% / 32% | 123% / 43% | 96% / 38% | 78% / 50% | 514% / 42% | 396% / 83% | 4% / 66% | 4% / 65% |
| Real SRAD | 3% / | 1% / | 0% / | 6% / | 165% / | 166% / | 47% / 561% | 47% / 431% |
| Virtual SRAD | 30% / 81% | 46% / 87% | 15% / 76% | 10% / 464% | 270% / 116% | 247% / 106% | 4% / 0% | 4% / 0% |
| Real B+Tree | 135% / 98% | 46% / 141% | 70% / 62% | 65% / 75% | 317% / 88% | 310% / 93% | 12% / 6% | 16% / 5% |
| Virtual B+Tree | 88% / 50% | 107% / 170% | 65% / 58% | 61% / 78% | 400% / 47% | 331% / 44% | 12% / 8% | 5% / 5% |

Figure 28: Impact Caused by Algorithms Implemented with the OpenCL or OpenMP Libraries.

Analysis of the results shows that all combinations of SRAD implemented in a real environment with OpenCL libraries competing with algorithms implemented with OpenMP libraries should be avoided. We noticed that the smallest loss suffered by this algorithm was 431% competing with B+Tree running in a virtual environment, and the largest loss was 836% competing with SRAD in a virtual environment. All combinations of SRAD in real or virtual environments implemented with the OpenMP library competing with algorithms implemented with OpenCL libraries should be avoided because the smallest loss suffered with this algorithm was 166% competing with SRAD running in a real environment, and the largest loss was 514% competing with Kmeans in a virtual environment.

The algorithms associated with the smallest loss were B+Tree in a real or virtual environment implemented with OpenMP libraries running concurrently with SRAD implemented with the OpenCL libraries and running in a virtual environment (without loss to the SRAD algorithm and with a loss of 4% for B+Tree algorithms in real or virtual environments). B+Tree implemented with OpenMP libraries running in a virtual environment concurrently with B+Tree implemented with OpenCL libraries running in a virtual environment showed a loss of 5% for the two algorithms, and LUD implemented with OpenCL libraries running concurrently in a virtual environment with B+Tree implemented with Open MP libraries running in a real environment showed a loss of between 8% and 4%.

Other combinations can coexist in accordance with the policies of access and control of the environment, mainly by evaluating the stability of the environment (less distance between points) and quality (lower average loss). It is incumbent upon the environment administrator or the scheduling algorithm to decide the best combination based on the data generated.

# 6 Conclusion

This paper presents an analysis of the effects of concurrency between classes of algorithms used in applications running in virtual and real environments. In this analysis, we observed the diverse effects caused by these combinations, which are associated with the libraries used in the implementation of the algorithms. The test results and the proposed model for evaluating these effects using the concept of affinity and through the adoption of a scale for its measurement show that this information could become a valuable resource for the analysis of the particular behavior of an application in a shared computational environment, the conditions of this type of environment or the allocation of new running applications or virtual environments in clouds.

This study is important because the use of virtual environments has become increasingly common. Virtual environments are being used as a way to optimize the use of real resources (consolidation) or when cloud computing is presented as a solution for obtaining additional resources on demand. The results obtained provide insight into the best combinations between the various applications that share these resources. The results of the experiments showed that, if there is real need for resource sharing, there is consequent concurrency for the physical resource. Some types of applications can coexist without significant degradation of the environment, allowing this sharing, while other combinations of applications should be avoided. As an extension of this work, the results and conclusions presented here can be used for developing applications and schedulers for the tested environments to minimize the performance loss resulting from the concurrency.

Figure 28 presents a summary of all the results obtained in the tests. The values presented consolidate all losses in the combinations tested, indicating the types of concurrency that can coexist on the same resource and which concurrencies should be avoided. With the data available from these tables, a scheduler task can choose the best combinations. The normalized scoring model resource is also considered an innovation, because until now, no scoring mechanism has been found that allows an evaluation of these algorithms and languages. The scheduler task seeks to equalize these values of different orders of magnitude, as is the case for the execution times of the algorithms. The scheduler task evaluates the entire set of algorithms tested through maximum and minimum scores in a pre-defined way, allowing a quantitative analysis of the impacts of concurrency. The scoring model creates performance metrics that are able to rate existing computing infrastructure to check its performance for the applications allocated to it, either in real or virtual environments. These metrics can be used to ease the use of this environment based on the previous knowledge about the effects of the concurrency, and in this way it is possible to delimit the choice of the best environment to run the needed algorithm concurrently, choosing between the best response time (less influence of another algorithm) or greater stability (when the response time is not critical), for example.

The evaluation of the average performance loss when there is heterogeneous competition proved to be a good approach to define combinations of libraries that can compete for the same resource and which libraries cannot compete. In the same review, by calculating the distance between points, the combinations of environment, algorithm, and library could provide better stability to the environment, defining two types of affinity, the first related to a quantitative criterion (the average performance loss of the combinations) and the second related to a qualitative criterion (the stability of the environment).

In heterogeneous testing using OpenMP and OpenCL, for all pairwise algorithm combinations, the combination of the types of library used caused significant variation in the performance of the environment. In this study, we could base the value of the loss of performance in each combination on the distance between the elements of the combination.

As a proposal for future work, continuation of the testing with all other algorithms of the Rodinia package is suggested. Furthermore, with the data obtained so far (from more than 7000 tests), it is possible to consolidate the results and extract relevant information for decision making to be used in scheduler implementations.

# Acknowledgments

This work has support from the National Research Council (CNPq) and the Brazilian Ministry of Science and Technology (MCT).

# References

[1] Upendra Sharma, Prashant Shenoy, Sambit Sahu, e Anees Shaikh. A cost-aware elasticity provisioning system for the cloud. *Proceedings of the 2011 31st International Conference on Distributed Computing Systems*, ICDCS '11, pages 559–570, Washington, DC, USA, 2011. IEEE Computer Society.
[2] CA Technologies. *The Complete Guide To Monitoring Virtualized Environments*. CA Technologies, 2013.
[3] Rodrigo N. Calheiros, Rajiv Ranjan, Anton Beloglazov, Cesar A. F. De Rose and Rajkumar Buyya, Cloudsim: a toolkit for modeling and simulation of cloud computing environments and evaluation of resource provisioning algorithms. *Softw. Pract. Exper.*, 41(1):23–50, janeiro de 2011.
[4] Kate Keahey, Patrick Armstrong, John Bresnahan, David LaBissoniere, e Pierre Riteau. Infrastructure outsourcing in multi-cloud environment. *Proceedings of the Workshop on Cloud Services, Federation, and the 8th Open Cirrus Summit*, San Jose, CA (USA), 2012.
[5] Vignesh T. Ravi, Michela Becchi, Gagan Agrawal, e Srimat Chakradhar. Supporting gpu sharing in cloud environments with a transparent runtime consolidation framework. *Proceedings of the 20th international symposium on High performance distributed computing*, HPDC '11, pages 217–228, New York, NY, USA, 2011. ACM.
[6] Phillip Colella. Defining software requirements for scientific computing. DARPA HPCS presentation, 2004.
[7] Erich L. Kaltofen. The seven dwarfs of symbolic computation*, 2010.
[8] Krste Asanovic, Ras Bodik, Bryan Christopher Catanzaro, Joseph James Gebis, Parry Husbands, Kurt Keutzer, David A. Patterson, William Lester Plishker, John Shalf, Samuel Webb Williams, e Katherine A. Yelick. The Landscape of Parallel Computing Research: A View from Berkeley. UCB/EECS-2006-183, EECS Department, University of California, Berkeley, Dec de 2006.
[9] Kevin Skadron. Rodinia:accelerating compute-intensive applications with accelerators, 2014. [Online; accessed 02-June-2014].
[10] Shuai Che, Michael Boyer, Jiayuan Meng, David Tarjan, Jeremy W. Sheaffer, Sang-Ha Lee, e Kevin Skadron. Rodinia: A benchmark suite for heterogeneous computing. *IISWC*, pages 44–54. IEEE, 2009.
[11] Laion F. Manfroi, Mariza Ferro, Andre Yokoyama, Antonio Roberto Mury, and Bruno Schulze. A walking dwarf on the clouds. *IEEE/ACM 6th International Conference on Utility and Cloud Computing*, pages 399–404, Dresden, Germany, 2013. International Workshop on Clouds and (eScience) Applications Management (CloudAM2013), IEEE Computer Society.
[12] A.S. Tanenbaum. *Modern Operating Systems*. GOAL Series. Prentice Hall, 2008.
[13] Christiane Pousa Ribeiro, Jean-Francois Mehaut, Alexandre Carissimi, Marcio Castro, e Luiz Gustavo Fernandes. Memory affinity for hierarchical shared memory multiprocessors. *Proceedings of the 2009 21st International Symposium on Computer Architecture and High Performance Computing*, SBAC-PAD '09, pages 59–66, Washington, DC, USA, 2009. IEEE Computer Society.
[14] Vegard Engen, Juri Papay, Stephen C. Phillips, e Michael Boniface. Predicting application performance for multi-vendor clouds using dwarf benchmarks. *Proceedings of the 13th international conference on Web Information Systems Engineering*, WISE'12, pages 659–665, Berlin, Heidelberg, 2012. Springer-Verlag.
[15] Luis M. Vaquero, Luis RoderoMerino, e Rajkumar Buyya. Dynamically scaling applications in the cloud. *SIGCOMM Comput. Commun. Rev.*, 41(1):45–52.